\documentclass[aps,pra,10pt,twocolumn,showpacs,superscriptaddress,footnoteinbib]{revtex4-1}
\usepackage[latin1]{inputenc}
\usepackage{amsmath}
\usepackage{amsfonts}
\usepackage{amssymb}
\usepackage{makeidx}
\usepackage{graphicx}
\usepackage{braket}
\usepackage{bm}
\usepackage{bbm}
\usepackage{color}
\usepackage{hyperref}
\usepackage{dsfont}
\usepackage{enumerate} 
\usepackage{float} 
\usepackage{mathrsfs}
\usepackage{subfigure}

\begin{document}

\title{Preservation of quantum coherence under Lorentz boost for narrow uncertainty wave packets}

\author{Riddhi Chatterjee}
\email{riddhi.chatterjee@bose.res.in}
\affiliation{S. N. Bose National Centre for Basic Sciences, Block JD, Sector III, Salt Lake, Kolkata 700098, India}

\author{A. S. Majumdar}
\email{archan@bose.res.in}
\affiliation{S. N. Bose National Centre for Basic Sciences, Block JD, Sector III, Salt Lake, Kolkata 700098, India}

\begin{abstract}
We consider the effect of relativistic boosts on single particle Gaussian wave packets. The coherence
of the wave function as measured by the boosted observer is studied as a function of the momentum and
the boost parameter. Using various formulations of coherence it is shown that in general the coherence decays with the increase of the momentum of the state, as well as the boost applied to it. Employing
a basis-independent formulation, we show however, that coherence may be preserved even for large boosts applied on narrow uncertainty wave packets. Our result is exemplified quantitatively for
practically realizable neutron wave functions.  
\end{abstract}


\maketitle

\section{Introduction}

The realization that the physical world is both relativistic and quantum mechanical at the
fundamental level has inspired the development of much of modern physics. Quantum information
science that has origins in some key foundational questions~\cite{epr,schrod,bell} raised in the previous century, has undergone a rapid phase of development over the last several years. However,
an overwhelming majority of such studies have been performed in the domain of nonrelativistic
quantum information. A number of information theoretic protocols though rely for their implementation
on photons for which there exists no nonrelativistic approximation.  

The relevance and impact of relativistic effects on the concepts of quantum information
was first pointed out by Peres \textit{et al.}~\cite{peres,peresrev}. In particular, considering a single qubit state in the framework of relativity, it was shown that the spin entropy of the qubit increases with respect to an inertial observer even due to pure boost as a result  of the coupling of the momentum degrees of freedom with the spin. The situation becomes worse in case of an arbitrary Lorentz transformation which may completely decohere a single qubit state forbidding single qubit communication without shared reference frames~\cite{17,18}. 

The study of relativistic quantum information is important not only due to the intricacies of
the fundamental issues involved, but also due to its applications in diverse domains as discussed
in several works. The relativistic generalization of the EPR experiment was first considered by
Czachor~\cite{czachor}. The effects of observer dependence on entanglement have been widely
studied by Fuentes \textit{et al.}~\cite{fuentes,fuentes_inrt}. It has been observed that relativistic considerations impose additional constraints on the security of quantum key distribution~\cite{barrett}. Additionally, relativistic quantum information is essential to the 
study of the black hole information paradox~\cite{bh}, and may be of relevance in information
theoretic concepts applied to quantum gravity~\cite{qg} and cosmology~\cite{cosmo}.   

It has been recently realized that quantum coherence~\cite{3} is the most basic feature of
quantumness of single systems responsible for superposition of quantum states, from which all
quantum correlations arise in composite systems. Defined in a quantitative manner based on the framework of resource theory~\cite{1},\cite{2},\cite{6,7,8,9,10,11,12}  quantum coherence may be exploited to perform quantum tasks. Several operational measures of quantum coherence have been proposed~\cite{4},\cite{5}, 
enabling it to be used for detection of genuine non-classicality in physical states. However,
as is the case with entanglement, there exists no unique quantifier of coherence. Problems 
of physical consistency arising
out of basis dependent formulations of coherence measures have been noted~\cite{13}.  On the
other hand, basis independent measures of coherence have also been formulated~\cite{14},\cite{15},
 which manifest intrinsic randomness contained in a quantum state.

In  the present work our motivation is to investigate the behaviour of quantum  coherence in the
relativistic scenario. Though relativistic quantum information has been studied earlier
in the context of entropies of single systems as well as entanglement of composite systems~\cite{peres,28,ent}, 
the question as to how coherence behaves under relativistic transformations remains to be analysed.
Our aim here is to partially fill this gap in the literature in the context of single particle states.   
Specifically, we  study quantitatively the change in coherence of a single particle Gaussian state under the application of Lorentz boosts employing various coherence quantifiers. Our results exhibit
a generic loss of coherence for the relativistic observer. However, using a basis independent measure we show that coherence may be preserved to a large extent for narrow wave packets enabling the possibility of single qubit communication  without sharing of reference frames. 

The plan of this paper is as follows. In the next section we present a brief overview of the different
basis dependent measures and one basis independent measure that we have used in our subsequent analysis. In
section III we provide a description of the behaviour of a single particle quantum state under
relativistic boost. In section IV we compute the coherence of a spin-$1/2$ particle with Gaussian
momentum distribution using the different measures of coherence. A specific example of a narrow
uncertainty wave packet using neutron parameters is presented in section V  showing that basis independent coherence is indeed preserved under relativistic boosts. 
We make some concluding
remarks in section VI.

\section{Mathematical preliminaries: coherence measures}

According to\textsuperscript{}, The defining properties that any functional $C$ mapping states $\rho$ to non-negative real numbers should satisfy in order for it to be a proper coherence measure, are~\cite{3}, (i) $C(\rho)$ should vanish for any incoherent state, (ii) monotonicity under incoherent
completely positive and trace preserving (ICPTP) maps, and (iii) convexity. 
Several candidate measures have been suggested which satisfy above criteria:
\paragraph*{$l_{1}$-norm:}
\begin{equation} \label{l1norm}
C_{l_{1}} = \sum_{\substack{i,j \\ i \neq j}} \vert \rho_{ij} \vert
\end{equation}
\paragraph*{Relative Entropy of Coherence:}
\begin{equation} \label{relen}
C_{rel. ent.}(\rho) = S(\rho_{diag}) - S(\rho)
\end{equation}
where $S$ is the von-Neumann entropy and $\rho_{diag}$ is the state containing only diagonal elements of $\rho$. 
\paragraph*{Skew Information:}  If an observable $X$ is measured on the state $\rho$, the skew information is given by~\cite{4},\cite{19},
\begin{equation}
\mathcal{I}(\rho, X) = -\dfrac{1}{2} Tr\lbrace[\sqrt{\rho},X]^{2}\rbrace
\end{equation}
Because of the square root term this quantity cannot be expressed in terms of observable but it is possible to set a nontrivial lower bound which can be measured experimentally. 
For a generic state of the form 
\begin{equation}
\rho = \dfrac{1}{2}(\mathds{1} + \vec{n}\cdot \mathbf{\Sigma})
\end{equation}
the skew information corresponding to the observable $\Sigma_{3}$ is given by\cite{19}
\begin{equation}\label{skew}
\mathcal{I}(\rho,\Sigma_{3}) = (1-\sqrt{1-\vert\vec{n}\vert^{2}}) (n_{1}^{2} + n_{2}^{2})
\end{equation}
where $\vec{n}$ is the bloch vector and $\lbrace \Sigma_{i} \rbrace$ are the Pauli matrices.

\subsection*{Basis Independent Measure of Coherence:}
The coherence quantifiers defined above are basis dependent, i.e, the amount of coherence in a quantum state quantified by those measures depends upon the bases in which the state is represented.  Recently, a basis independent quantifier of coherence has been defined, which measures the intrinsic randomness contained in a quantum state. A Frobenius-norm based measure~\cite{14} is defined as 
\begin{equation}\label{fn}
\mathscr{C}(\rho) = \sqrt{\dfrac{d}{d-1}} \lVert \rho - \rho_{\star} \rVert_{F}
\end{equation}
where $d$ is the dimension of the Hilbert space that spans $\rho$, and $\rho_{\star} = \mathbb{I}_{d}/d$ is the maximally mixed state. The Frobenius-norm is given by $\lVert A \rVert_{F} = \sqrt{Tr\left(A^{\dagger}A\right)}$. Frobenius-norm is normalized to guarantee $\mathscr{C}(\rho) \in [0,1]$. The most significant property of this measure is that it is basis independent, i.e, unitary invariant, $\mathscr{C}(\rho) = \mathscr{C}(U \rho U^{\dagger})$ owing to the fact that the maximally mixed state $\rho_{\star}$ is the only state that remains invariant under arbitrary unitary transformations. 
Eq.(\ref{fn}) can be rewritten as 
\begin{equation}\label{fn1}
\mathscr{C}(\rho) = \sqrt{\dfrac{d}{d-1}\sum^{d}_{j=1}\left(\lambda_{j} - \dfrac{1}{d}\right)^{2}}
\end{equation}
where $\lbrace\lambda_{j}\rbrace$ is the eigenvalues of $\rho$. The above quantity  is a measure of purity, and  $\mathscr{C}^{2}(\rho)$ is proportional to the Brukner-Zeilinger information (BZI)~\cite{20} which is an operational notion defined as the sum of individual measures of information over a complete set of mutually complementary observables (MCO)~\cite{21}. BZI is itself invariant under the unitary transformation of the quantum state or equivalently of the choice of the measured set of MCO. 

\section{Single particle quantum state under relativistic boost:}       

In Minkowski space-time positive energy, massive, single particle states furnish a spinor representation of the Poincar\'{e} group~\cite{fuentes_inrt,24,26}. The bases  of representation space are labelled by $\lbrace \ket{\mathbf{p},j} \rbrace$, where $\mathbf{p}$ is the spatial components of 
the $4$-momentum $p^{\mu}$ with $p^{0}=\sqrt{\mathbf{p}^{2}+m^{2}}$. $m$ is the rest mass of the particle. $j$ is total angular momentum along a quantization axis and equal to the intrinsic spin $s$ of the particle in its rest frame. Normalization is defined as~\cite{24}
$\braket{\mathbf{p}',j'\mid \mathbf{p},j} = \bm{\delta}(\mathbf{p}'-\mathbf{p}) \ \delta_{j'j}$.
For Lorentz transformation $\Lambda$ the basis state transforms under unitary transformation $U(\Lambda)$ given by
\begin{equation}
U(\Lambda)\ket{\mathbf{p},j} = \sqrt{\dfrac{(\Lambda p)^{0}}{p^{0}}} \sum_{j'}D_{j j'}(W(\Lambda,\mathbf{p}))\ket{\Lambda \mathbf{p},j'}
\end{equation}
We will assume $j$ to be discrete. $\Lambda \textbf{p}$ is the spatial component of the Lorentz transformed $4$-momentum. $W(\Lambda,\mathbf{p})$ is an element of the little group of the Poincar\'{e} group and $D(W(\Lambda,\mathbf{p}))$ is its unitary representation. 

For a massive particle $W(\Lambda,\mathbf{p}) \in SO(3)$,  hence $D(W(\Lambda,\mathbf{p})) \in SU(2)$.
 If the $4$-momentum of the particle  is parametrized by
\begin{equation}
p^{\mu} = (m \cosh{\beta}, m \sinh{\beta}\hat{f})
\end{equation}
where $m$ be the mass of the particle, and the velocity of the frame $O^{\Lambda}$ is $\mathbf{v} = \tanh{\alpha}\ \mathbf{\hat{e}}$ then,  the representation of $D(W(\Lambda,\mathbf{p}))$  is
given by\cite{26,30}
\begin{equation}\label{repr1}
D(W(\Lambda,\mathbf{p})) = \cos{\dfrac{\phi}{2}} \mathds{1} + i \sin{\dfrac{\phi}{2}} (\mathbf{\Sigma}\cdot \mathbf{\hat{n}})
\end{equation}
where 
\begin{equation}\label{repr2}
\cos{\dfrac{\phi}{2}} = \dfrac{\cosh{\dfrac{\alpha}{2}}\cosh{\dfrac{\beta}{2}} + \sinh{\dfrac{\alpha}{2}}\sinh{\dfrac{\beta}{2}}(\mathbf{\hat{e}\cdot \hat{f}})}{\sqrt{\dfrac{1}{2} + \dfrac{1}{2} \cosh{\alpha}\cosh{\beta} + \dfrac{1}{2} \sinh{\alpha} \sinh{\beta}(\mathbf{\hat{e}\cdot \hat{f}})}}
\end{equation}
\begin{equation}\label{repr3}
\sin{\dfrac{\phi}{2}} \mathbf{\hat{n}} = \dfrac{\sinh{\dfrac{\alpha}{2}}\sinh{\dfrac{\beta}{2}}(\mathbf{\hat{e}\times \hat{f}})}{\sqrt{\dfrac{1}{2} + \dfrac{1}{2} \cosh{\alpha}\cosh{\beta} + \dfrac{1}{2} \sinh{\alpha} \sinh{\beta}(\mathbf{\hat{e}\cdot \hat{f}})}}
\end{equation}
with $\phi$ and $\mathbf{\hat{n}}$ being respectively, the  angle and axis of Wigner rotation.

A pure state may be written as 
\begin{equation}\label{sep}
\ket{\psi} = \sum_{s} \int d\mathbf{p}{\ } \psi(\mathbf{p}) \ket{\mathbf{p}} \otimes a_{s} \ket{s}
\end{equation}
w.r.t laboratory reference frame $O$. An observer $O^{\Lambda}$ boosted by Lorentz transformation $\Lambda$ w.r.t $O$ sees the state (\ref{sep}) as 
\begin{equation}\label{spmoent}
\ket{\psi^{\Lambda}} = \sum_{s} \int d\mathbf{p}{\ } \sqrt{\dfrac{(\Lambda p)^{0}}{p^{0}}} \psi(\mathbf{p})\  a_{s} \sum_{s'} D_{s s'}(W(\Lambda,\mathbf{p}))\ket{\Lambda \mathbf{p},s'}
\end{equation}
The state in equation  (\ref{sep}) is separable in spin and momentum, but not the state in equation (\ref{spmoent}), since $W(\Lambda,\mathbf{p})$ is a function of momentum $p$ and so is $D(W(\Lambda,\mathbf{p}))$. In equation (\ref{spmoent}) the basis states have undergone a momentum dependent rotation known as Wigner rotation resulting in the coupling between spin and momentum which is known as spin-momentum entanglement~\cite{peres,27}.

A single particle spin-$1/2$ state~\cite{fuentes_inrt,27} given by 
\begin{equation}
\rho = \sum_{s_{1},s_{2}} \int \int d\mathbf{p}_{1} d\mathbf{p}_{2} {\ } \psi(\mathbf{p}_{1})\psi^{\ast}(\mathbf{p}_{2}) {\ } a_{s_{1}} a^{\ast}_{s_{2}} \ket{\mathbf{p}_{1},s_{1}}\bra{\mathbf{p}_{2},s_{2}}
\end{equation}
 may be traced over the momentum degrees of freedom to obtain the spin reduced density matrix, given by~\cite{26}
\begin{equation}
\begin{split}
\rho_{s} & = \sum_{s_{1},s_{2}} \int \int \int d\mathbf{p} d\mathbf{p}_{1} d\mathbf{p}_{2} {\ } \psi(\mathbf{p}_{1})\psi^{\ast}(\mathbf{p}_{2}) \\
& \qquad \qquad \qquad a_{s_{1}} a^{\ast}_{s_{2}} \braket{\mathbf{p} \mid \mathbf{p}_{1},s_{1}}\braket{\mathbf{p}_{2},s_{2} \mid \mathbf{p}} \\ 
& = \sum_{s_{1},s_{2}} \int \int d\mathbf{p} {\ } \psi(\mathbf{p})\psi^{\ast}(\mathbf{p}) {\ } a_{s_{1}} a^{\ast}_{s_{2}} \ket{s_{1}}\bra{s_{2}}
\end{split}
\end{equation}
The  density matrix in the frame of boosted observer is given by
\begin{equation}
\begin{split}
\rho^{\Lambda} & = \sum_{s_{1},s_{2}} \int \int d\mathbf{p}_{1}d\mathbf{p}_{2}{\ } \sqrt{\dfrac{(\Lambda p_{1})^{0}{\ }(\Lambda p_{2})^{0}}{p_{1}^{0}{\ }p_{2}^{0}}}{\ } \psi(\mathbf{p}_{1}) \psi^{\ast} 
(\mathbf{p}_{2}) a_{s_{1}} a^{\ast}_{s_{2}}\\ 
& \sum_{s_{1}', s_{2}'} D_{s_{1} s_{1}'}(W(\Lambda,\mathbf{p}_{1}))\ket{\Lambda \mathbf{p}_{1},s_{1}'}\bra{\Lambda \mathbf{p}_{2},s_{2}'} D^{\dagger}_{s_{2} s_{2}'}(W(\Lambda,\mathbf{p}_{2}))
\end{split}
\end{equation}
The corresponding reduced density matrix is hence given by
\begin{equation}\label{sud_cal}
\begin{split}
& \rho^{\Lambda}_{s} = 
\sum_{\substack{s_{1},s_{2}, \\ s_{1}', s_{2}'}} \int d\mathbf{p}\ \lvert \psi(\mathbf{p}) \rvert^{2} \\ 
& \hspace*{2 cm} D_{s_{1} s_{1}'}(W(\Lambda,\mathbf{p}))\ket{s_{1}'}\bra{s_{2}'} D^{\dagger}_{s_{2} s_{2}'}(W(\Lambda,\mathbf{p}))
\end{split}
\end{equation}
where we have used 
$\delta(\Lambda \mathbf{p}_{1} - \Lambda \mathbf{p}_{2}) = \dfrac{(p_{1})^{0}}{(\Lambda p_{1})^{0}}\delta(\mathbf{p}_{1} - \mathbf{p}_{2})$. 
The reduced density matrix defined in this way is not covariant, as the transformation law of 
the secondary variable (spin) depends not only upon the Lorentz transformation $\Lambda$, but also upon the primary variable (spatial component of the $4$-momentum).  A boosted single particle Gaussian wave packet of the form $e^{-\mathbf{p}^{2}/2\sigma^{2}}$ was studied in Ref.\cite{peres} to obtain the von-Neumann entropy of the spin reduced density matrix (SRDM) in both the rest and the Lorentz boosted frames. A larger entropy was obtained in the boosted frame $O^{\Lambda}$  indicating the loss of information. The entropies corresponding to SRDM have been also obtained in other works assuming $4$-momentum to be discrete~\cite{28,27,29}.

\section{Coherence of a Spin-1/2 particle with Gaussian momentum distribution under relativistic boost}

Let us consider the single particle state 
\begin{equation}
\ket{\psi} = \dfrac{1}{\sqrt{2}} \int d\mathbf{p}{\ } \psi(\mathbf{p}) \ket{\mathbf{p}} \otimes (\ket{0} + \ket{1})
\end{equation}
with momentum $p^{\mu} = (m \cosh{\beta}, m \sinh{\beta}\hat{x}) = (p^{0},p_{x}\hat{x})$ w.r.t to the observer $O$ (for simplicity we consider the one dimensional velocity of the particle). 
The density matrix corresponding to the state $\ket{\psi}$ is given by
\begin{equation}
\rho = \dfrac{1}{2} \int\int d\mathbf{p}_{1}d\mathbf{p}_{2}{\ } \psi(\mathbf{p}_{1})\psi^{\ast}(\mathbf{p}_{2})\ket{\mathbf{p}_{1}}\bra{\mathbf{p}_{2}}\otimes (\mathds{1} + \sigma_{1})
\end{equation}
Assuming $\psi(\mathbf{p})$ to be the normalised SRDM corresponding to $\rho$, one has
\begin{equation}
\rho_{s} = \dfrac{1}{2}(\mathds{1} + \sigma_{1})
\end{equation}

In the frame of $O^{\Lambda}$ moving with velocity $v = \tanh{\alpha}\ \hat{z}$, the state of the particle is given by
\begin{equation}\label{blm}
\ket{\psi^{\Lambda}} = \dfrac{1}{\sqrt{2}} \int d\mathbf{p}{\ } \sqrt{\dfrac{(\Lambda p)^{0}}{p^{0}}} \psi(\mathbf{p}) D(W(\Lambda,\mathbf{p})) (\ket{\Lambda \mathbf{p},0} + \ket{\Lambda \mathbf{p},1})
\end{equation}
where 
\begin{equation}\label{D1}
D(W(\Lambda,\mathbf{p})) = \cos{\dfrac{\phi_{p_{x}}}{2}} + i \sin{\dfrac{\phi_{p_{x}}}{2}}\sigma_{2}
\end{equation}
 and
\begin{equation}
\cos{\dfrac{\phi_{p_{x}}}{2}} = \dfrac{\cosh{\dfrac{\alpha}{2}}\cosh{\dfrac{\beta}{2}} }{\sqrt{\dfrac{1}{2} + \dfrac{1}{2} \cosh{\alpha}\cosh{\beta}}} \nonumber
\end{equation}
\begin{equation}\label{D2}
\sin{\dfrac{\phi_{p_{x}}}{2}} = \dfrac{\sinh{\dfrac{\alpha}{2}}\sinh{\dfrac{\beta}{2}}}{\sqrt{\dfrac{1}{2} + \dfrac{1}{2} \cosh{\alpha}\cosh{\beta}}}
\end{equation}
with the axis of rotation being along the direction $\hat{z} \times \hat{x} = \hat{y}$. Substituting Eq.(\ref{D1},\ \ref{D2}) in Eq.(\ref{blm}) we have 
\begin{equation}
\begin{split}
& \ket{\psi^{\Lambda}} = \dfrac{1}{\sqrt{2}} \int d\mathbf{p}{\ } \sqrt{\dfrac{(\Lambda p)^{0}}{p^{0}}} \psi(\mathbf{p}) \Big[(\cos{\dfrac{\phi_{p_{x}}}{2}} + \sin{\dfrac{\phi_{p_{x}}}{2}})\ket{\Lambda \mathbf{p},0}\\
& \hspace*{3.7 cm} + (\cos{\dfrac{\phi_{p_{x}}}{2}} - \sin{\dfrac{\phi_{p_{x}}}{2}})\ket{\Lambda \mathbf{p},1}\Big]
\end{split}
\end{equation}
The density matrix corresponding to the state $\ket{\psi^{\Lambda}}$ is given by
\begin{equation}
\begin{split}
&\rho^{\Lambda} = \dfrac{1}{2} \int \int d\mathbf{p}_{1}d\mathbf{p}_{2}{\ } \sqrt{\dfrac{(\Lambda p_{1})^{0}{\ }(\Lambda p_{2})^{0}}{p_{1}^{0}{\ }p_{2}^{0}}} \psi(\mathbf{p}_{1})\psi^{\ast}(\mathbf{p}_{2}) \\
& \Big[ A_{p_{x1}}A_{p_{x2}}\ket{\Lambda \mathbf{p}_{1},0}\bra{\Lambda \mathbf{p}_{2},0}+ A_{p_{x1}}B_{p_{x2}}\ket{\Lambda \mathbf{p}_{1},0}\bra{\Lambda \mathbf{p}_{2},1} + \\ & A_{p_{x2}}B_{p_{x1}}\ket{\Lambda \mathbf{p}_{1},1}\bra{\Lambda \mathbf{p}_{2},0} + B_{p_{x1}}B_{p_{x2}}\ket{\Lambda \mathbf{p}_{1},1}\bra{\Lambda \mathbf{p}_{2},1} \Big]
\end{split}
\end{equation}
where $A_{p_{xi}} = (\cos{\dfrac{\phi_{p_{xi}}}{2}} + \sin{\dfrac{\phi_{p_{xi}}}{2}})$ and $B_{p_{xi}} =
(\cos{\dfrac{\phi_{p_{xi}}}{2}} - \sin{\dfrac{\phi_{p_{xi}}}{2}})$. 
Using Eq.(\ref{sud_cal}) the SRDM corresponding to $\rho^{\Lambda}$ is given by
\begin{equation}\label{srdm}
\begin{split}
\rho^{\Lambda}_{s} = \dfrac{1}{2} & \int d\mathbf{p} \  \left|\psi(\mathbf{p})\right|^{2} \Big[ A_{p_{x}}^{2}\ket{0}\bra{0} \\
& + A_{p_{x}}B_{p_{x}}\big(\ket{0}\bra{1} + \ket{1}\bra{0}\big) + B_{p_{x}}^{2}\ket{1}\bra{1}\Big]
\end{split}
\end{equation}

We will now  calculate $\rho^{\Lambda}_{s}$ for two particular forms of $\psi(\mathbf{p})$. Since we have assumed the velocity of the particle to be along $x$-axis, we will consider the following forms of $\psi(\mathbf{p}) = f(p_{x})\delta(p_{y})\delta(p_{z})$ with $f(p_{x})$ given by
\begin{enumerate}[(i)]
\item{\textbf{case (i):} (corresponding to the Gaussian wave packet centred at zero) $f(p_{x})=\dfrac{1}{(\sqrt{\pi}\sigma)^{1/2}}\ e^{-\dfrac{1}{2}\Big(\dfrac{p_{x}}{\sigma}\Big)^{2}}$}
\item{\textbf{case (ii):} (corresponding to the Gaussian wave packet centred at $\mathfrak{p}$) $f(p_{x})=\dfrac{1}{(\sqrt{\pi}\sigma)^{1/2}}\ e^{-\dfrac{1}{2}\Big(\dfrac{p_{x}-\mathfrak{p}}{\sigma}\Big)^{2}}$, where $\mathfrak{p}$ is a constant.}
\end{enumerate}
Eq.(\ref{srdm}) may be hence written as 
\begin{equation}
\begin{split}
\rho^{\Lambda}_{s} = \dfrac{1}{2} & \int dp_{x} \  \left|f(p_{x})\right|^{2} \Big[ A_{p_{x}}^{2}\ket{0}\bra{0} \\ 
& + A_{p_{x}}B_{p_{x}}\big(\ket{0}\bra{1} + \ket{1}\bra{0}\big) + B_{p_{x}}^{2}\ket{1}\bra{1}\Big]
\end{split}
\end{equation}
Henceforth we will use $p$ instead of $p_{x}$ for convenience. Now substituting $\cosh{\beta} = \sqrt{1 + \dfrac{p^{2}}{m^{2}}}$ , $\sinh{\beta} = \dfrac{p}{m}$, $\cosh{\alpha} = b$ and $\sinh{\alpha} = a$ we get 
\begin{equation}
\begin{split}
& A_{p}^{2} = 1 + \dfrac{a\  \dfrac{p}{m}}{1+b\sqrt{1 + \dfrac{p^{2}}{m^{2}}}} \\
& B_{p}^{2} = 1 - \dfrac{a\  \dfrac{p}{m}}{1+b\sqrt{1 + \dfrac{p^{2}}{m^{2}}}} \\
& A_{p}B_{p}  = \dfrac{b + \sqrt{1 + \dfrac{p^{2}}{m^{2}}}}{1+b\sqrt{1 + \dfrac{p^{2}}{m^{2}}}}
\end{split}
\end{equation} 
The components of the SRDM $\rho^{\Lambda}_{s}$ are given by
\begin{equation} 
\begin{split}
& \rho^{\Lambda}_{s\ 11} = \dfrac{1}{2} \int dp \  \left|f(p)\right|^{2} \left(1 + \dfrac{a\  \dfrac{p}{m}}{1+b\sqrt{1 + \dfrac{p^{2}}{m^{2}}}}\right) \\
& \rho^{\Lambda}_{s\ 22} = \dfrac{1}{2} \int dp \  \left|f(p)\right|^{2} \left(1 - \dfrac{a\  \dfrac{p}{m}}{1+b\sqrt{1 + \dfrac{p^{2}}{m^{2}}}}\right) \\
& \rho^{\Lambda}_{s\ 12} = \rho^{\Lambda}_{s\ 21} =\dfrac{1}{2} \int dp \  \left|f(p)\right|^{2} \left(\dfrac{b + \sqrt{1 + \dfrac{p^{2}}{m^{2}}}}{1+b\sqrt{1 + \dfrac{p^{2}}{m^{2}}}}\right)
\end{split}
\end{equation}
Under the approximation $\left(\dfrac{\sigma}{m}\right)\ll1$, we obtain the components of the density matrix analytically, given by

%%%%%%%%%%%%%%%%%%%%%%%%%%%%%%%%%%%%%

\begin{figure*}[t]
\begin{center}
\begin{minipage}[h]{.4\linewidth}
\centering 
   \includegraphics[width=\columnwidth]{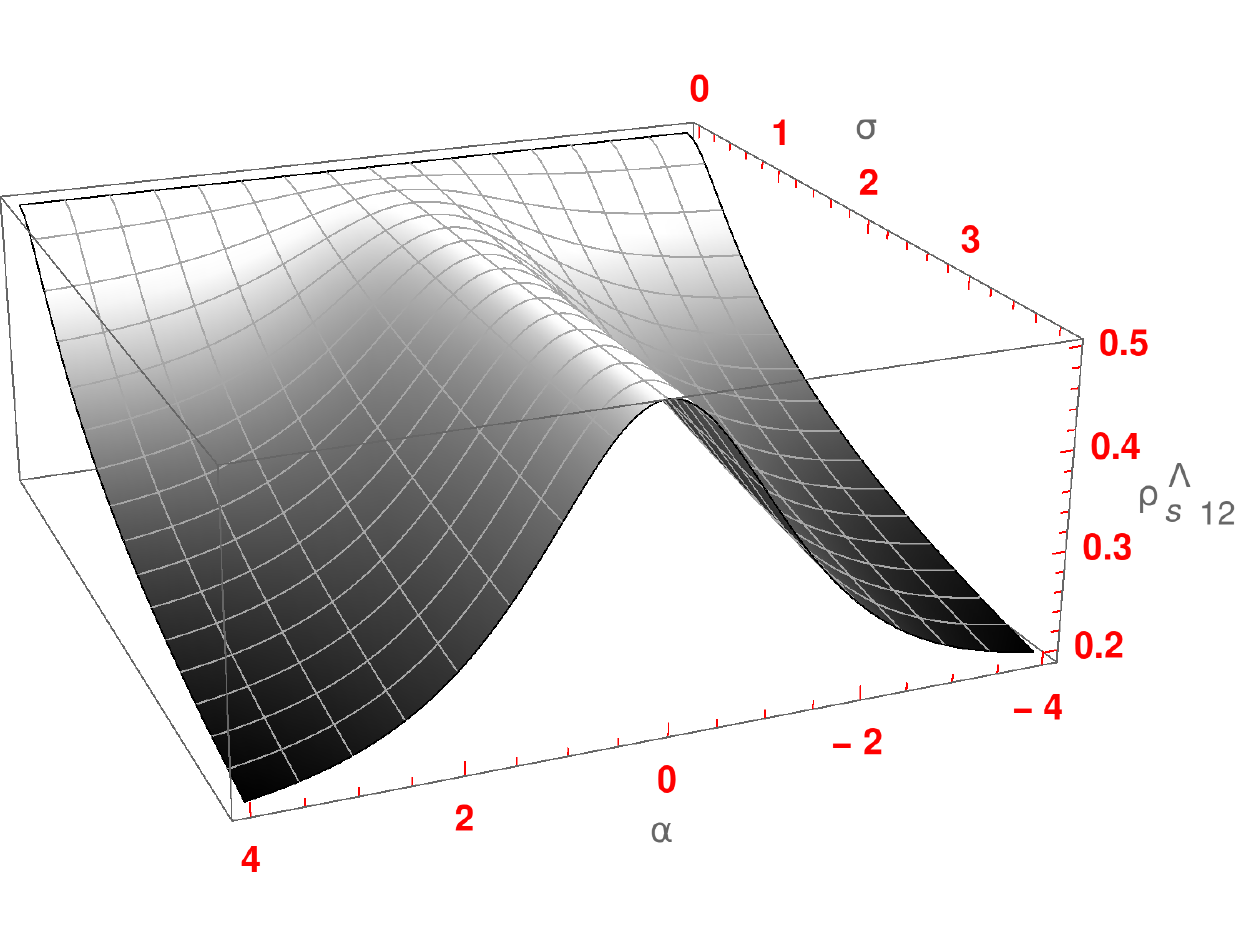}
        \caption{\small $\rho^{\Lambda}_{s\ 12}$ vs. $\alpha$ and $\sigma$(MeV) for wave packet centred at zero.}
        \label{l0_rho12n}
\end{minipage}
    \hspace{2.5cm}
\begin{minipage}[h]{.4\linewidth}
\centering
   \includegraphics[width=\columnwidth]{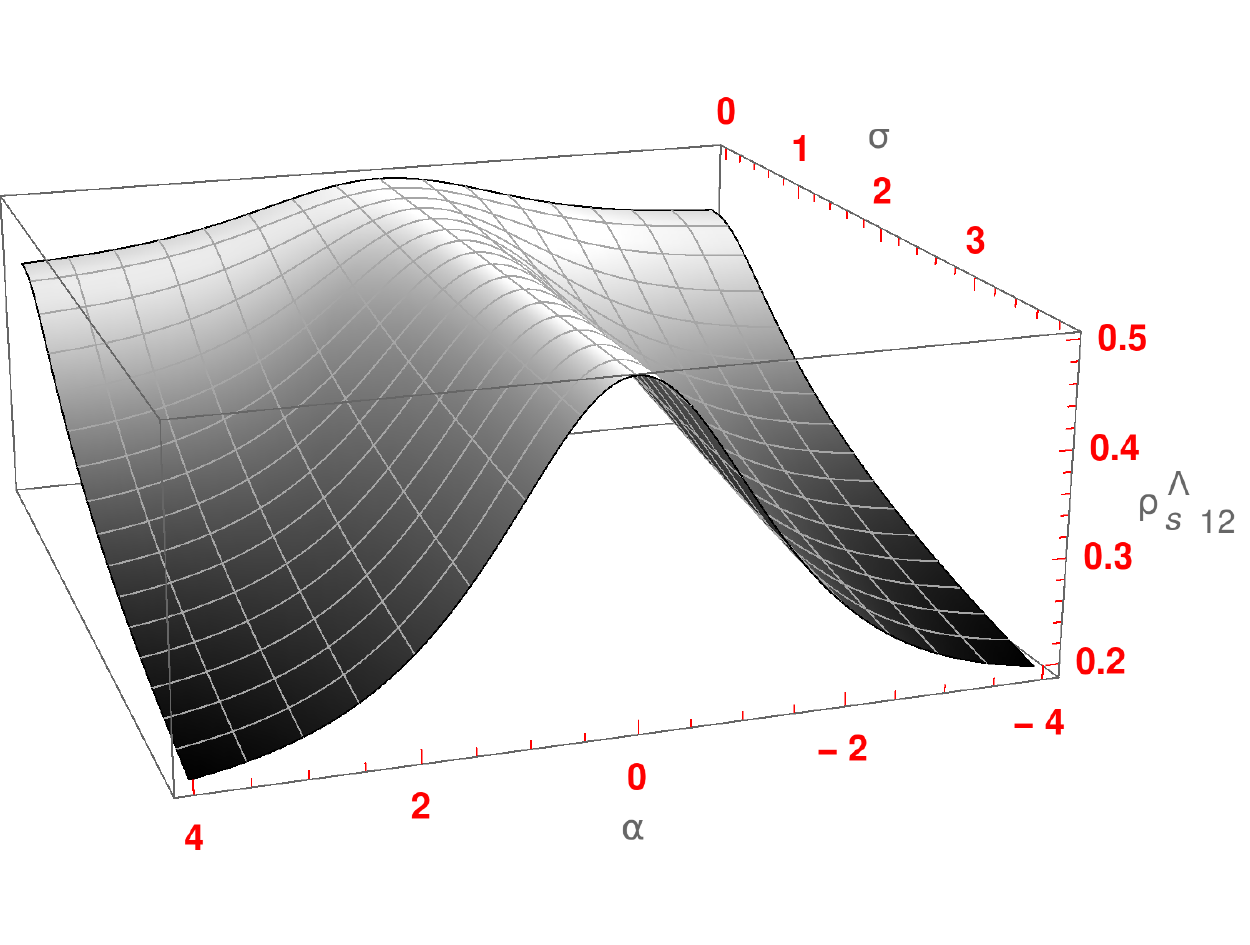}
        \caption{\small $\rho^{\Lambda}_{s\ 12}$ vs. $\alpha$ and $\sigma$(MeV) for wave packet centred at $\mathfrak{p}$.}
        \label{lp_rho12n}
\end{minipage}
\end{center}
\end{figure*}

%%%%%%%%%%%%%%%%%%%%%%%%%%%%%%%%%%%%%%%

\begin{equation}\label{analytic}
\begin{split}
& \rho^{\Lambda}_{s\ 11} = \rho^{\Lambda}_{s\ 22} = \dfrac{1}{2} \\
& \rho^{\Lambda}_{s\ 12} = \rho^{\Lambda}_{s\ 21} = \dfrac{1}{2} - \dfrac{1}{8} \left(\dfrac{\cosh{\alpha}-1}{\cosh{\alpha}+1}\right) \left(\dfrac{\sigma}{m}\right)^{2} 
\end{split}
\end{equation}

For larger uncertainty we calculate the integral numerically. We first plot the dependence of $\rho^{\Lambda}_{s\ 12} = \rho^{\Lambda}_{s\ 21}$ with respect to uncertainty $\sigma$ of the state and the rapidity parameter $\alpha$  for boosted observer in the
figures \ref{l0_rho12n} and \ref{lp_rho12n}. The purpose of studying $\rho^{\Lambda}_{s\ 12} = \rho^{\Lambda}_{s\ 21}$ is that in the basis dependent framework the coherence is manifested by
the  off-diagonal elements of the density matrix. So, a decrease in $\rho^{\Lambda}_{s\ 12}$ implies decoherence. This results from  spin momentum entanglement induced by the Wigner rotation~\cite{peres}.
The computations displayed in the plots are done with taking mass $m \approx 0.5$ MeV (case of an electron) and $\mathfrak{p} = 1/2\sqrt{3}$ MeV (momentum of electron moving with half of the speed of light).  It can be checked using Eq.(\ref{analytic}) that for small values of $\left(\dfrac{\sigma}{m}\right)$ the analytical result provides a good approximation to the  numerical calculation. As $\sigma \rightarrow 0$, it can be observed that $\rho^{\Lambda}_{s\ 12} \rightarrow 1/2$. The amount of this decoherence  increases with $\alpha$ as seen in the plots. In case of the wave packet centred at zero (Figure  \ref{l0_rho12n} ), there is no decoherence as $\sigma \rightarrow 0$, whatever is the value of $\alpha$. This is because zero uncertainty  implies that the particle is at rest and the pure boost due to $O^\Lambda$  does not induce Wigner rotation. However,  for $\sigma \neq 0$ there is decoherence due to spin-momentum entanglement. In case of the wave packet centred at $\mathfrak{p}$ (Figure \ref{lp_rho12n}), decoherence due to spin momentum entanglement is again clearly exhibited. Note that there is Wigner rotation  due to two noncolinear boosts corresponding to the motion of particle and the observer respectively.
 When the uncertainty tends to zero, i.e.,   $\sigma \rightarrow 0$, implying that the  momentum of the particle tends to a single sharp value $\mathfrak{p}$,  there is no spin-momentum entanglement.  But since  the quantum state will undergo a pure rotation $(\cos{\dfrac{\phi_{p}}{2}}\mathds{1} + i \sin{\dfrac{\phi_{p}}{2}} \Sigma_{2})$ in this case, the basis dependent density matrix elements
 undergo a corresponding change. This is evident from the drop in the values of $\rho^{\Lambda}_{s\ 12}$ with increasing $\alpha$ even for $\sigma \rightarrow 0$ in Figure \ref{lp_rho12n}. \par
 
\begin{center}
\begin{figure*}[t]
\begin{minipage}[h]{.45\linewidth}
   \includegraphics[width=\columnwidth]{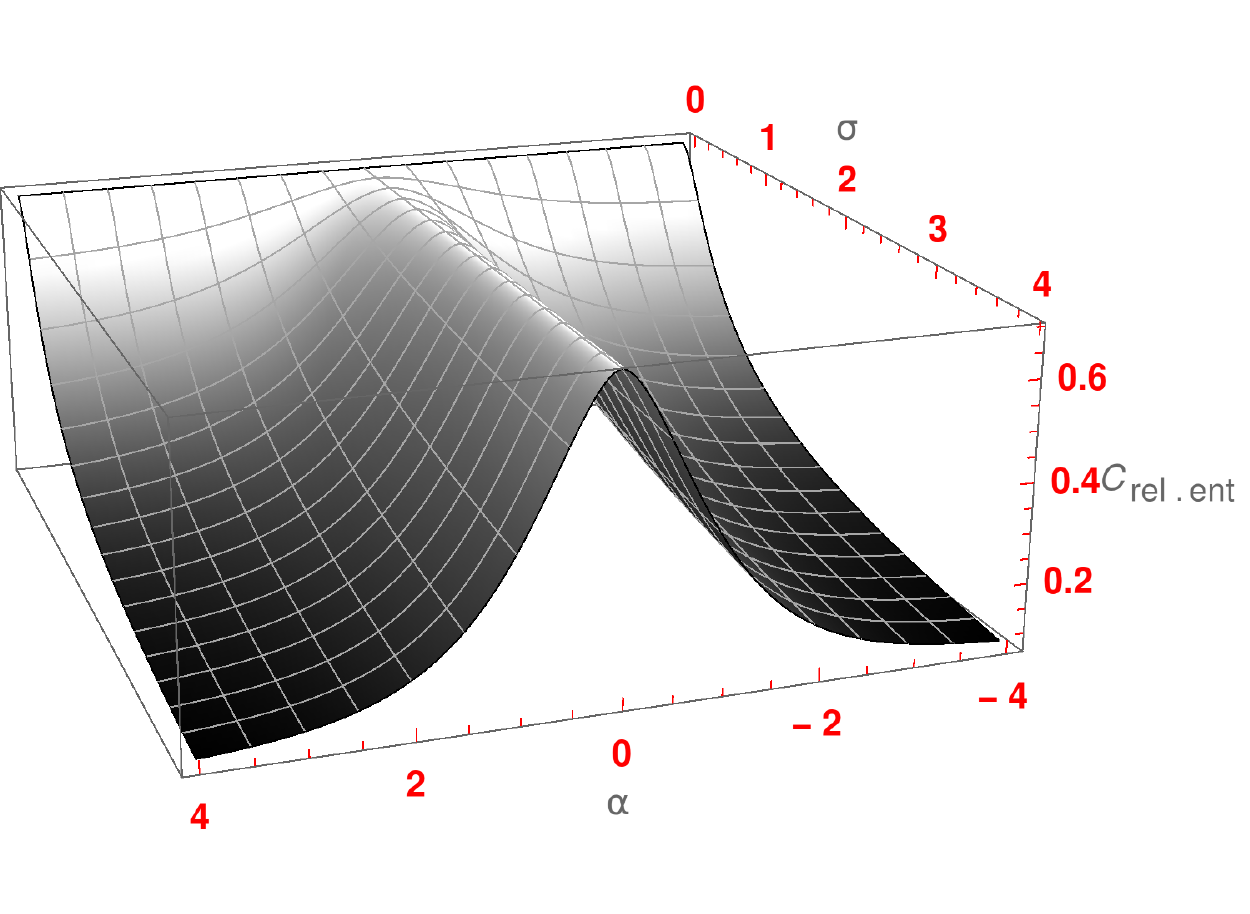}
        \caption{\small $C_{rel.ent}$ vs. $\alpha$ and $\sigma$(MeV) for wave packet centred at zero.}
        \label{l0_ren}
\end{minipage}
    \hfill
\begin{minipage}[h]{.45\linewidth}
   \includegraphics[width=\columnwidth]{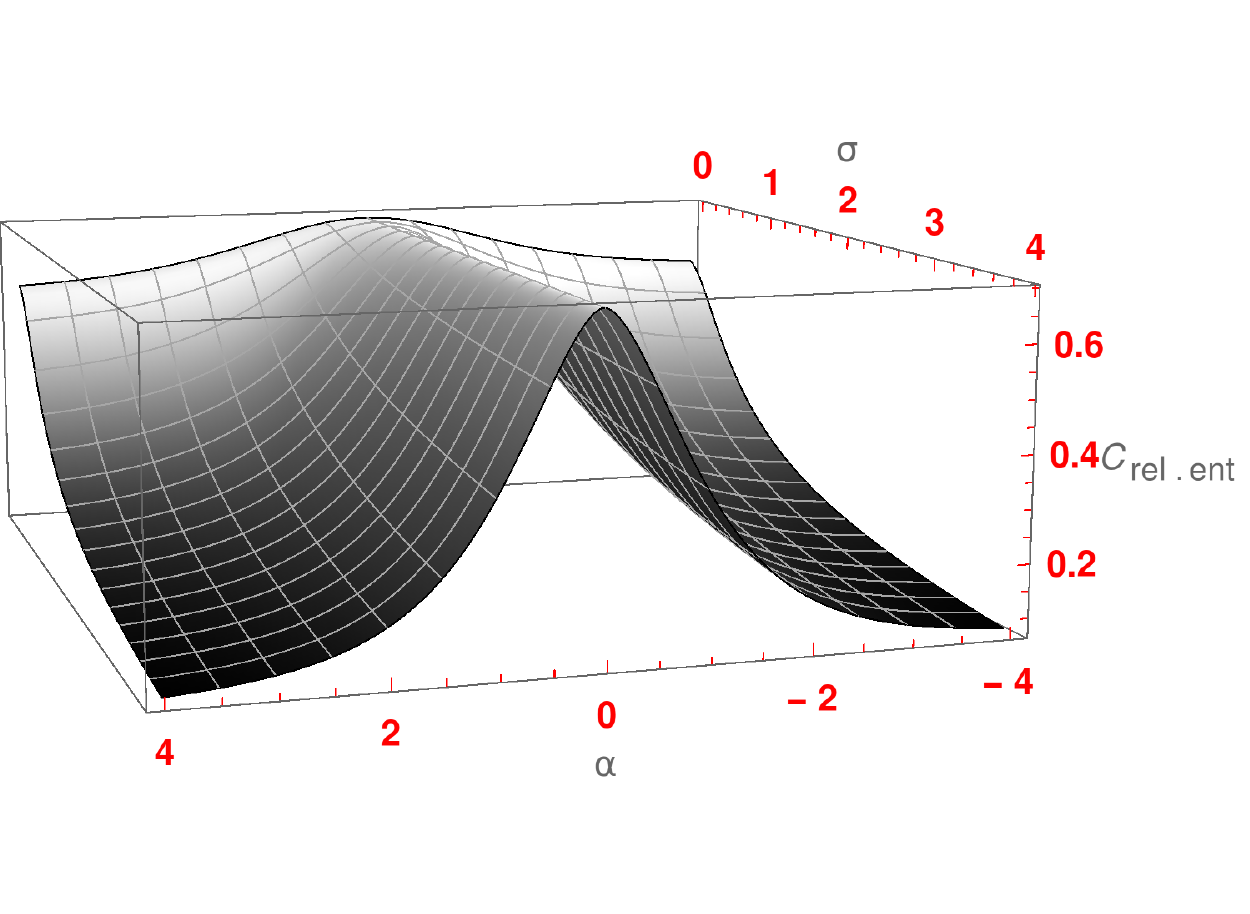}
        \caption{\small $C_{rel.ent}$ vs. $\alpha$ and $\sigma$(MeV) for wave packet centred at $\mathfrak{p}$.}
        \label{lp_ren}
\end{minipage}
    \hfill
\begin{minipage}[h]{.45\linewidth}
   \includegraphics[width=1.05\columnwidth]{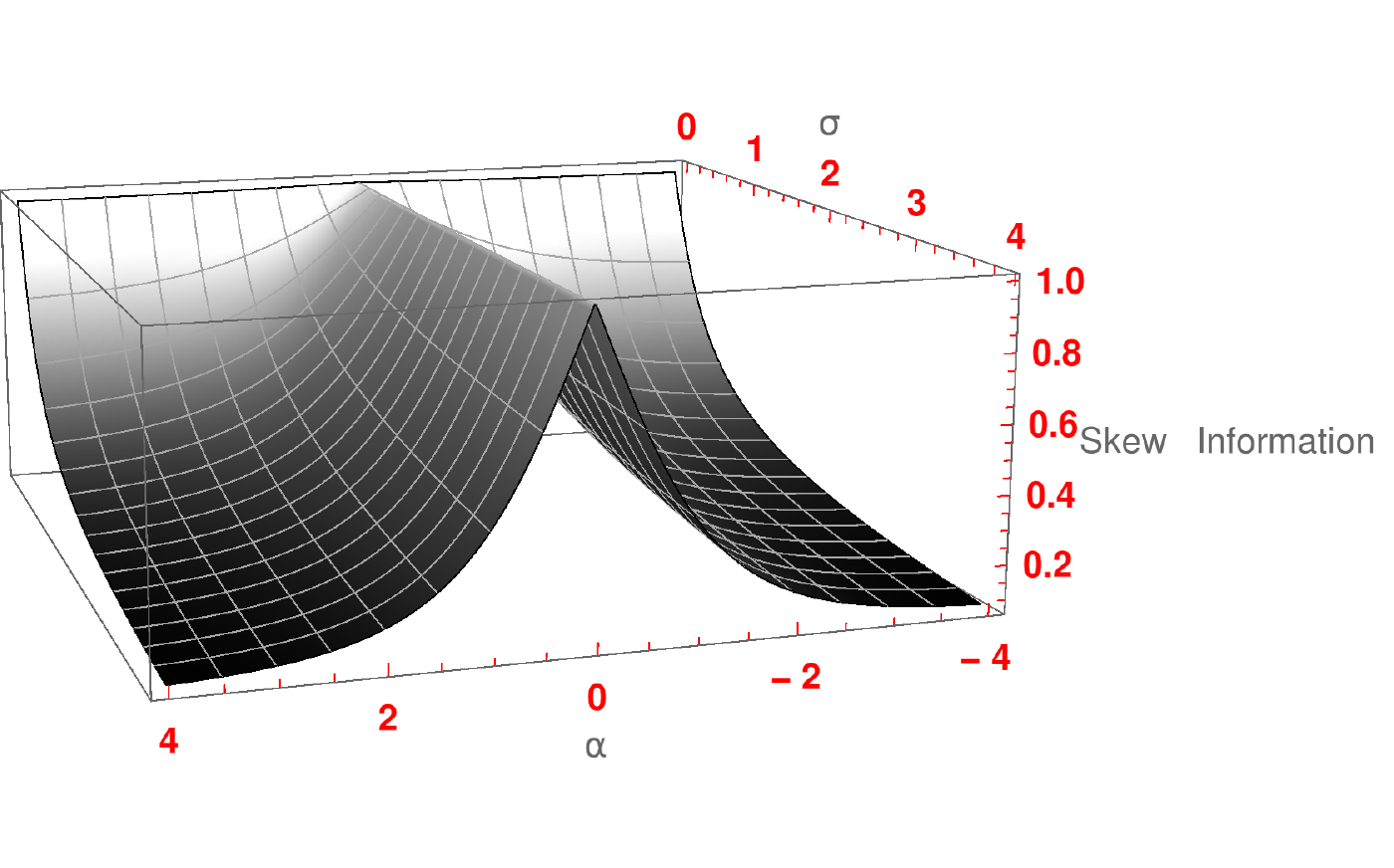}
        \caption{\small "Skew Information" $\mathcal{I}$ vs. $\alpha$ and $\sigma$(MeV) for wave packet centred at zero.}
        \label{l0_skn}
\end{minipage}
    \hfill
\begin{minipage}[h]{.45\linewidth}
   \includegraphics[width=\columnwidth]{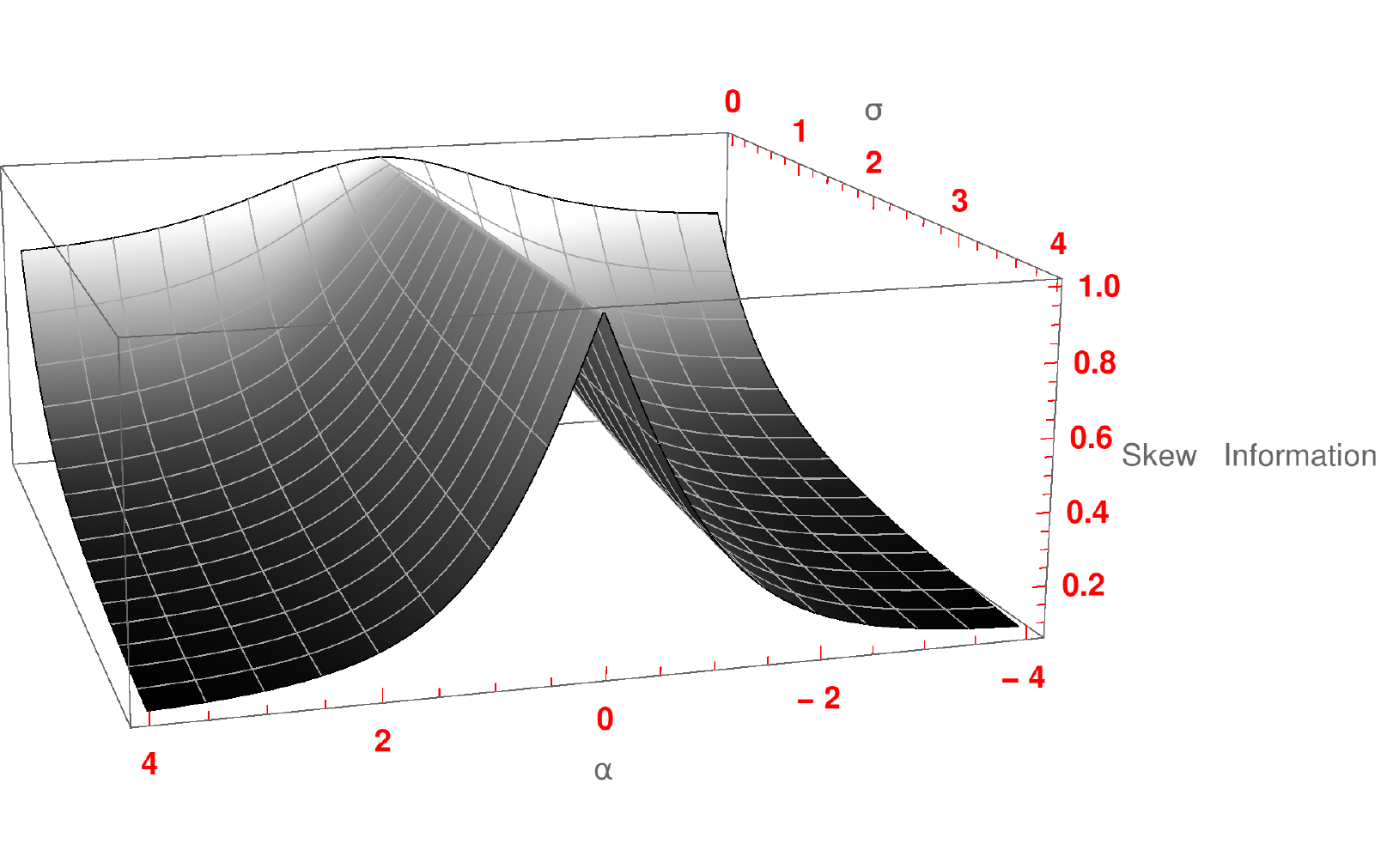}
        \caption{\small "Skew Information" $\mathcal{I}$ vs. $\alpha$ and $\sigma$(MeV) for wave packet centred at $\mathfrak{p}$.}
        \label{lp_skn}
\end{minipage}
\end{figure*}
\end{center}

Now with the components of $\rho^{\Lambda}_{s}$ we will study the change in coherence under relativistic boost using the different coherence quantifiers mentioned in Section II. First we study the basis dependent quantifiers. Note first that in this case the $l_{1}$-norm (\ref{l1norm}) is simply 
$C_{l_{1}} = 2\rho^{\Lambda}_{s\ 12}$, and hence, its values can be read off from the figures 
\ref{l0_rho12n} and \ref{lp_rho12n}.
The maximum value of the coherence of the state $\rho_{s}$ is $1$ measured by the $l_{1}$-norm and 
the skew information (\ref{skew}), and $\ln{2}$ when calculated with relative entropy (\ref{relen}). The coherence corresponding to
$\rho_{s}^{\Lambda}$ using the relative entropy measure is plotted in the figures \ref{l0_ren} and
\ref{lp_ren} for wave packets centered at zero and $\mathfrak{p}$, respectively. Similarly, the skew information versus the uncertainty and the boost parameter is plotted in the figures \ref{l0_skn} and \ref{lp_skn}\ . As discussed earlier it is clear from plots that in the case of wave packets centred at zero (figures \ref{l0_ren} and \ref{l0_skn}), the coherence is maximum if  either $\alpha$ or $\sigma$ goes to zero. For wave packets centred at $\mathfrak{p}$ (figures \ref{lp_ren} and \ref{lp_skn} ), the coherence attains its  maximum value only when $\alpha = 0$. The reason for the sharp edge in the plot of skew information is the positive square-root in the
equation (\ref{skew}). \par

\begin{center}
\begin{figure*}[t]
\begin{minipage}[h]{.45\linewidth}
   \includegraphics[width=\columnwidth]{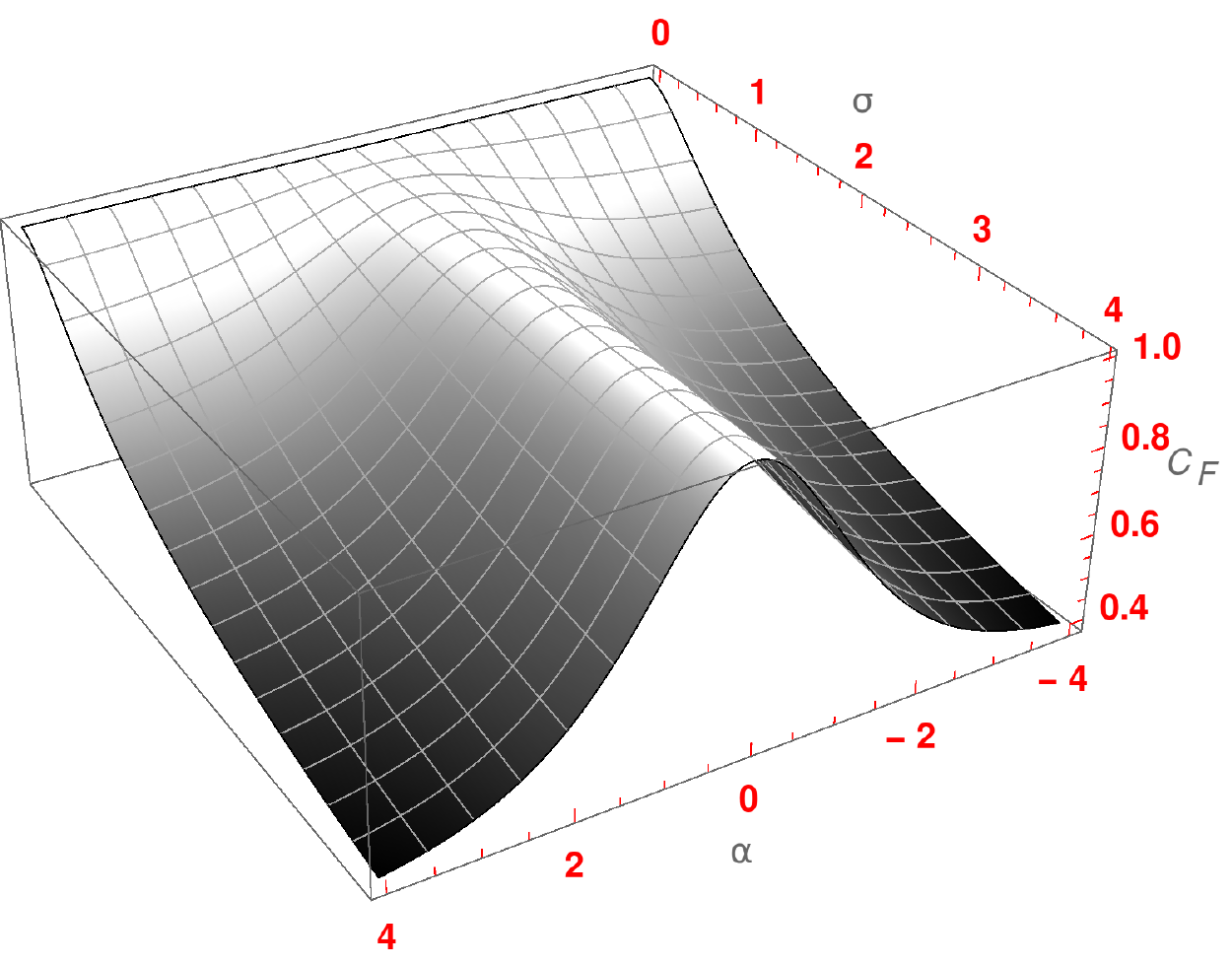}
        \caption{\small "Frobenius norm based measure of coherence"  $\mathscr{C}$ vs. $\alpha$ and $\sigma$(MeV) (for wave packet centred at zero.)}
        \label{l0_fn}
\end{minipage}
    \hfill
\begin{minipage}[h]{.45\linewidth}
   \includegraphics[width=\columnwidth]{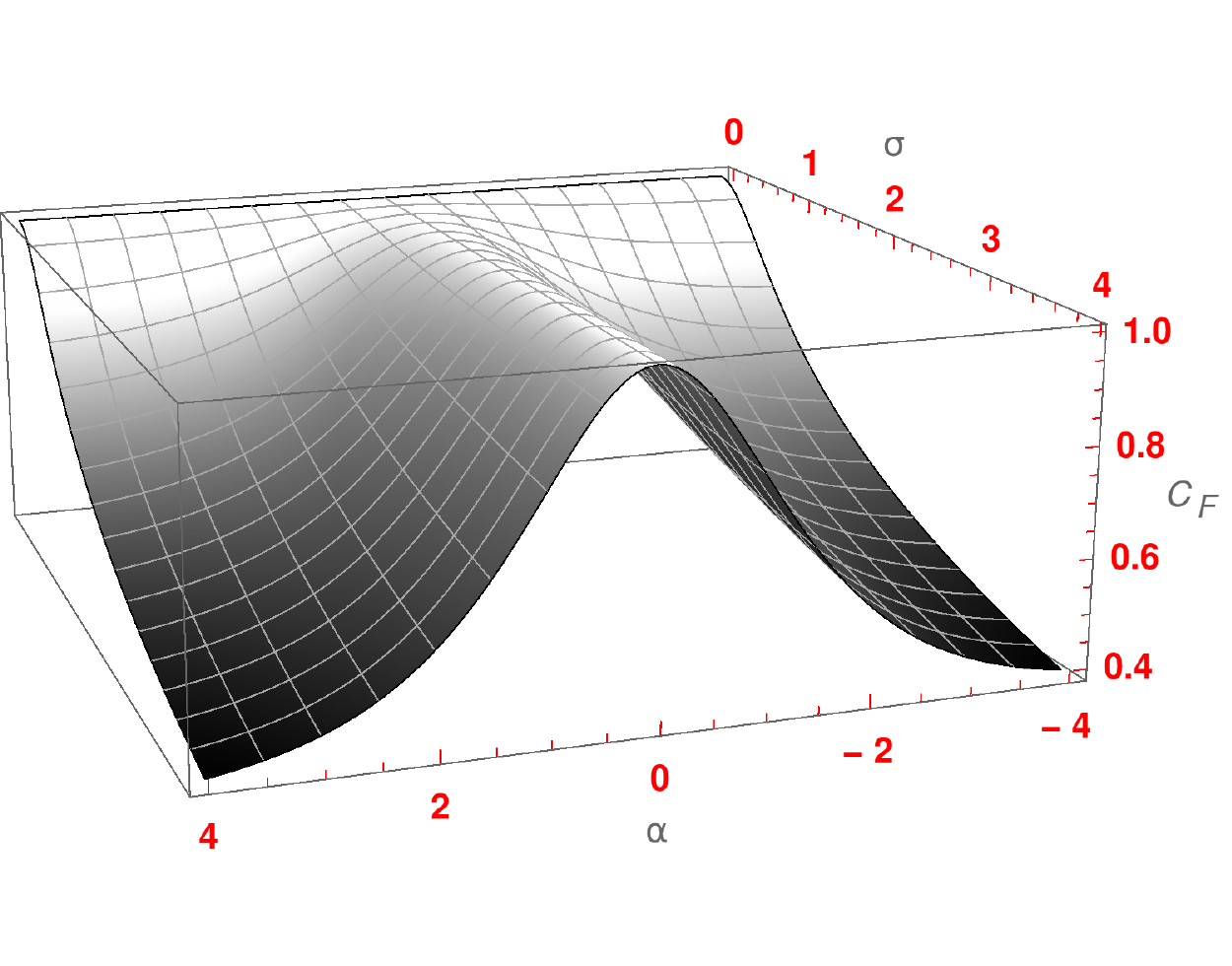}
        \caption{\small "Frobenius norm based measure of coherence" $\mathscr{C}$ vs. $\alpha$ and $\sigma$(MeV) (for wave packet centred at $\mathfrak{p}$.)}
        \label{lp_fn}
\end{minipage}
\end{figure*}
\end{center}

The above plots all correspond to  basis dependent measures of coherence. If we demand that coherence
should represent a physical property of the system  independent of the choice of bases, we should 
consider the coherence of a single particle quantum state under relativistic boost using a basis independent measure. Now, using the Frobenius norm based measure (\ref{fn}) we compute the coherence($C_{F}$) of the state $\rho_{s}$. For the state $\rho_{s}^{\Lambda}$. This is displayed in the  figures \ref{l0_fn} and  \ref{lp_fn}  for  \textit{case (i)} and \textit{case (ii)}, respectively. It can be seen from both the figures that the value of $C_{F}$ does not fall off with increasing boost $\alpha$ when $\sigma$ goes to zero. This is a significant result even when the wave packet is centered at $\mathfrak{p}$, contrasting with

 %%%%%%%%%%%%%%%%%%%%%%%%%%%%%%%%%%%%%%%%%%%%%%%%%%%%%%%%%%%%%%%%%%%%%%%%%%%%%%%%%%%%%%%%%%%%%%

the case of all of the basis dependent measures, i.e., $C_{l_1}$ from figure \ref{lp_rho12n}, $C_rel.ent$ from figure \ref{lp_ren}, and skew information $\cal{I}$ from figure \ref{lp_skn}.

The invariance of the Frobenius norm under unitary transformation~\cite{14} leads to preservation of this
basis independent measure of coherence for small uncertainty wave packets, since a pure basis
rotation is unable to impact the value of coherence even for large boosts. So, if there are two parties Alice ($O$) and Bob ($O^{\Lambda}$) who do not share any reference frame, and suppose Alice possesses a single party state whose spread in momentum is narrow enough. Then Bob, a relativistically moving observer can  access  the qubit from his own frame in which he will not see the qubit decohered.
It can be checked that
such a feature would also be obtained by using more general 3-dimensional wave packets. However,
when the value of $\left(\dfrac{\sigma}{m}\right)$ increases, decoherence becomes effective due to spin momentum entanglement for
large $\alpha$, as expected.

%%%%%%%%%%%%%%%%%%%%%%%%%%%%%%%%%%%%%%%%%%%%%%%%%%%%%%%%%%%%%%%%%%%%%%%%%%%%%%%%%%%%%%%%
%%%%%%%%%%%%%%%%%%%%%%%%%%%  up to here %%%%%%%%%%%%%%%%%%%%%%%%%%%%%%%%%%%%
%%%%%%%%%%%%%%%%%%%%%%%%%%%%%%%%%%%%%%%%%%%%%%%%%%%%%%%%%%%%%%%%%%%%%%%%%%%%%%%%%%%%

\section{Examples}
 
Let us now consider the specific case of a narrow uncertainty wave packet. For the situation in
which the particle is nonrelativistic in Alice's frame having low momentum, there exist several
 techniques to produce narrow uncertainty wave packets, such as using hydrogen atoms cooled in millikelvin range~\cite{18},  or with
 ultracold neutrons (UCN) having average kinetic energy $<$ 300 neV. UCN because of their low energy  are very sensitive to magnetic, gravitational and material potentials~\cite{34}. Their  significance
  in quantum gravity experiments have been proposed~\cite{35}, \cite{36}. It is regarded that a gravitational field would make a qubit decohere~\cite{37}, and hence,  it would be interesting to  apply the Frobenius norm based measure of coherence in examples involving the action of gravity
  on quantum states.
Below we provide a particular example of computation of the  Frobenius norm based measure of coherence using neutron parameters. 

\begin{center}
\begin{figure}[h]
\begin{minipage}[h]{\linewidth}
   \includegraphics[width=\columnwidth]{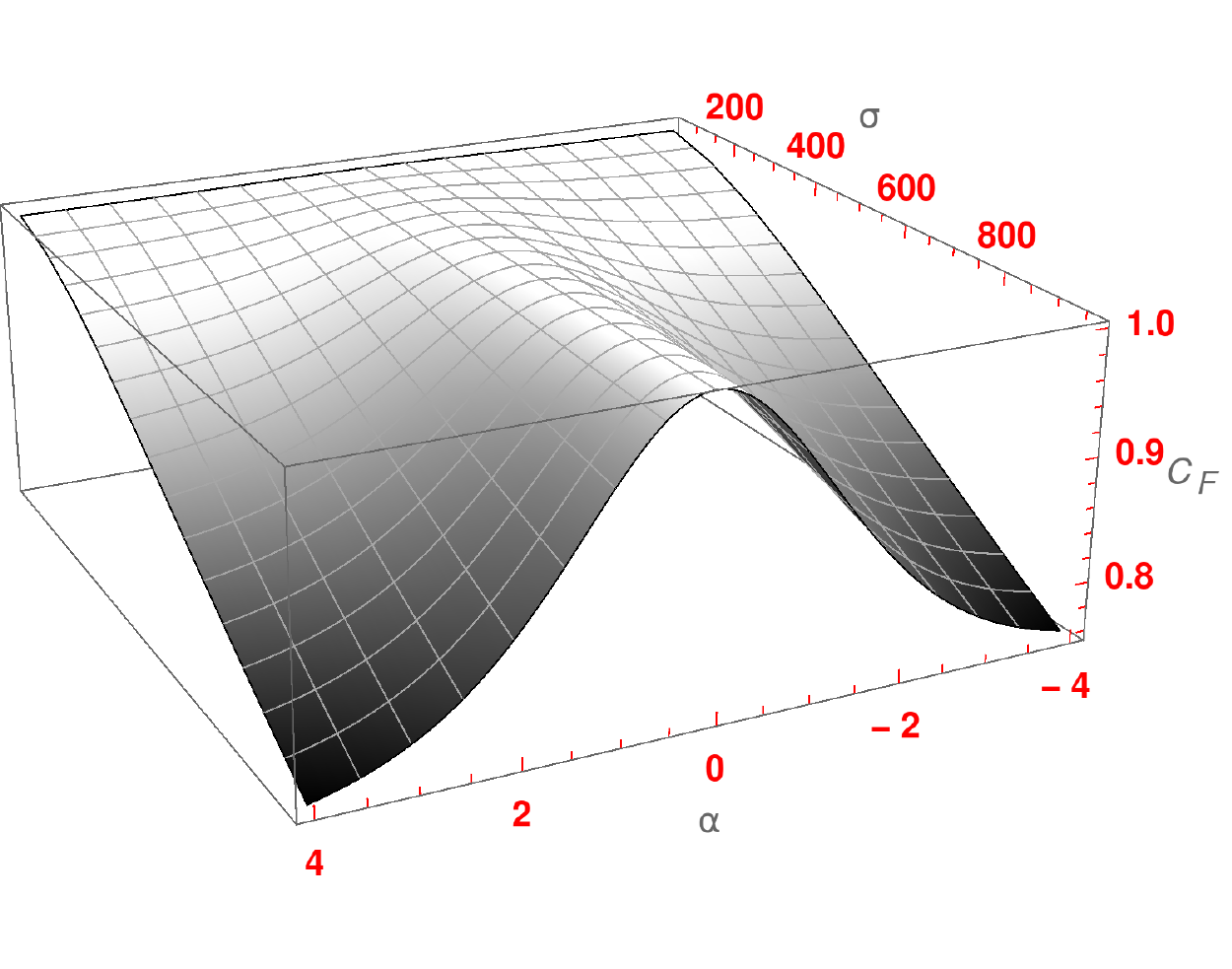}
        \caption{\small "Frobenius norm based measure of coherence" vs. $\alpha$ and $\sigma$(MeV) for
        narrow uncertainty neutron wave-packet}
        \label{l0_ucn}
\end{minipage}
\end{figure}
\end{center}

Let us consider the state 
\begin{equation}
\ket{\psi} = \dfrac{1}{\left(\sqrt{\pi}\sigma\right)^{3/2}}\int d\mathbf{p}\ e^{-\dfrac{\mathbf{p}^2}{2 \sigma^2}}\ket{p}\otimes\ket{0}
\end{equation}
with respect to frame $O$, where $\mathbf{p} = (p_{x},p_{y},p_{z})$ is the 3-momentum of the particle and $p^{0} = \sqrt{\mathbf{p}^2 + m^2}$. The boosted observer $O^\Lambda$ has velocity $v = \tanh{\alpha} \hat{z}$.
Thus, using the equations ( \ref{repr1} , \ref{repr2}\ , \ref{repr3}) we find the representation of Wigner's little group given by
\begin{equation}
\begin{split}
& D(W(\Lambda,\mathbf{p})) = \dfrac{1}{\left[ (p^0 + m) (p^0 \cosh{\alpha} + p_{z} \sinh{\alpha} + m) \right]^{1/2}} \times \\
& \left[ (p^0 + m)\cosh{\dfrac{\alpha}{2}} + p_{z}\sinh{\dfrac{\alpha}{2}} - i \sinh{\dfrac{\alpha}{2}}(-p_{x}\sigma_{y} + p_{y}\sigma_{x}) \right]
\end{split}
\end{equation}
From the above expression we can calculate transformed state $\ket{\psi^{\Lambda}}$. The SRDM corresponding to $\ket{\psi^{\Lambda}}$ is given by 
\begin{equation}
\rho^{\Lambda}_{s} = \dfrac{1}{\left(\sqrt{\pi}\sigma\right)^{3}}\int d\mathbf{p}\ e^{-\dfrac{\mathbf{p}^2}{\sigma^2}}\  \begin{pmatrix}
\dfrac{M}{A B} & 0\\
0 & \dfrac{N}{A B}
\end{pmatrix} \nonumber
\end{equation}
where 
\begin{equation}\label{neutron}
\begin{split}
& A = (p^0 + m)\\
& B = (p^0 \cosh{\alpha} + p_{z} \sinh{\alpha} + m)\\
& M = A^2 \cosh^{2}{\dfrac{\alpha}{2}} + p^2_{z}\sinh^{2}{\dfrac{\alpha}{2}} + A\  p_{z} \sinh{\alpha}\\
& N = (p^{2}_{x} + p^{2}_{y})\sinh^{2}{\dfrac{\alpha}{2}}
\end{split}
\end{equation}
Using equation (\ref{fn1}) we obtain the coherence of the state $\rho^{\Lambda}_{s}$ which is
displayed in the figure \ref{l0_ucn}, where we have used the rest mass of neutron $939.36$ MeV. It
can be seen that the loss of coherence is negligible even for large values of the boost $\alpha$ when $(\dfrac{\sigma}{m})$ is small.

%\onecolumngrid
%\twocolumngrid

For a narrow uncertainty wave packet the components of the Bloch vector can be obtained
analytically~\cite{peres} in the
 approximation $(\dfrac{\sigma}{m}) \ll 1$:
\begin{equation}
\rho^{\Lambda}_{s} = \dfrac{1}{2}\begin{pmatrix}
1+n_{z} & 0\\
0 & 1-n_{z}
\end{pmatrix} \nonumber
\end{equation}
where 
\begin{equation}
n_{z} = 1 - \left( \dfrac{\sigma}{2 m}\tanh{\dfrac{\alpha}{2}} \right)^2
\end{equation}
Using the above formula the Frobenius norm measure of coherence in this case turns
out to be
\begin{equation} \label{ucncoh}
\mathscr{C}(\rho^{\Lambda}_{s}) = (1 - \left( \dfrac{\sigma}{2 m}\tanh{\dfrac{\alpha}{2}} \right)^2).
\end{equation}
 In case of the UCN the upper bound of the kinetic energy is around $300 neV$. Assuming this value to represent the upper bound of $\sigma$, from Eq.(\ref{ucncoh})  we see that the loss of information
 due to decoherence resulting from relativistic spin-momentum entanglement is rather negligible of the order $\sim 10^{-30}$. 
There is  another kind of experimentally available neutron called the thermal neutron whose average kinetic energy is around $0.025 eV$~\cite{38}. With the corresponding  value of $sigma$ it can be
checked that the loss of coherence is again negligible of the order $\sim 10^{-20}$.

\section{Conclusions}

In the present work we have studied the behaviour of various coherence quantifiers under relativistic boosts. We find using several measures of coherence such as the $l_1$-norm, the relative entropy of coherence~\cite{3}, and the skew information~\cite{4}, that a relativistic observer measures a 
reduced value of coherence compared to the coherence of the pure quantum state in its rest frame.
Such a result follows as expected from the coupling of the the spin and momentum degrees of freedom that originates due to the
Wigner rotation encountered by the single particle quantum state under relativistic boost~\cite{peres}.
The above form of decoherence is a generic feature obtained for all measures of coherence, including
a basis independent formulation~\cite{14} that we have employed here. The most significant aspect
of our results is however, the preservation of coherence measured through the basis independent
Frobenius norm for the case of narrow uncertainty wave packets. We have shown explicitly using
neutron state parameters that the loss of coherence is negligible for not only ultracold but thermal
neutrons as well. This makes it possible for a relativistic observer to recognize a narrow uncertainty wave packet as a pure state with the help of the above measure.

Our analysis indicates that in order to place coherence as resource in relativistic framework, basis independent formulations are necessary. Recently, resource theory of asymmetry or reference frame  has emerged which treats individual formulations of coherence measures as special cases~\cite{6,7,8,9,10,11,12}. It has been noted that both operational and geometric perspectives are in general significant for resource theory~\cite{3}-\cite{2},\cite{14},\cite{23}.
Since Frobenius-norm has a geometric perspective~\cite{22}, the basis independent measure of coherence~\cite{14}  employed here clearly has geometric interpretation.  Moreover, the Frobenius norm based measure is square of the BZI and hence,  has an operational notion too. There are several protocols of communication using the BZI, such as in the case of quantum state estimation~\cite{31}, quantum teleportation~\cite{32}, and violation of Bell
inequalities~\cite{33}. Following the formulation~\cite{31}, it may be feasible in a relativistic framework for Bob to perform quantum state estimation of the qubit possessed by Alice. Communication using single partite states without sharing reference frames~\cite{17,18} may thus indeed be possible

{\it Acknowledgements:} ASM acknowledges support from the project SR/S2/LOP-
08/2013 of the DST, India.

\end{document}